\documentclass[amsmath,amssymb,prc,final,showpacs,twocolumn]{revtex4}
 
\usepackage{bm,hyphenat,xspace}
\usepackage{graphicx,epsfig}

\newcommand{\Fig}{Fig.}
\newcommand{\Figs}{Figs.}
\newcommand{\Ref}{Ref.}
\newcommand{\Refs}{Refs.}

\newcommand{\fm}{\;\mathrm{fm}}

\newcommand{\cm}{\mathrm{c\!\:\!.m\!\:\!.}}

\newcommand{\Hh}{{}^3\mathrm{H}}
\newcommand{\nH}{n\text{-}{}^3\mathrm{H}}
\newcommand{\pHe}{p\text{-}{}^3\mathrm{He}}

\newcommand{\nd}{n\text{-}d}

\newcommand{\vlowk}{$V_{\mathrm{low}\: k}\;$}
 
\begin{document}

\title{Low-momentum interactions in three- and four-nucleon scattering}
  
\author{A.~Deltuva} 
\email{deltuva@cii.fc.ul.pt}
\affiliation{Centro de F\'{\i}sica Nuclear da Universidade de Lisboa, 
P-1649-003 Lisboa, Portugal }

\author{A.~C.~Fonseca} 
\affiliation{Centro de F\'{\i}sica Nuclear da Universidade de Lisboa, 
P-1649-003 Lisboa, Portugal }

\author{S.~K.~Bogner} 
\affiliation{National Superconducting Cyclotron Laboratory and Department
of Physics and Astronomy, Michigan State University, East Lansing, MI 48824}

\received{December 7, 2007}

\begin{abstract}
Low momentum two-nucleon interactions  obtained with the renormalization group method 
and the similarity renormalization group method are used to study the cutoff dependence of low 
energy $3N$ and $4N$ scattering observables. The residual cutoff dependence arises from 
omitted short-ranged $3N$ (and higher) forces that are induced by the renormalization group 
transformations, and may help to estimate the sensitivity of various $3N$ and $4N$ scattering
observables to short-ranged many-body forces. 
\end{abstract}

 \maketitle


\section{Introduction}\label{Intr}

Modern few- and many-body calculations of nuclear structure and reactions are based on the picture of point-like nucleons interacting via two- and three-nucleon potentials.  For this purpose, a number of high precision $(\chi^2/\mathrm{datum} \simeq 1)$ but phenomenological meson exchange models of the two-nucleon ($2N$) force such as the Nijmegen~\cite{stoks:94a}, Argonne V18 (AV18) \cite{wiringa:95a} and CD-Bonn~\cite{machleidt:01a} potentials have been developed over the past decade. However, with phenomenological models it is not clear how to construct consistent three-nucleon ($3N$) forces and other operators. The lack of a systematic organization or counting scheme results in model-dependent predictions, as there is no way to make controlled comparisons between the different force models. More recently, substantial progress has been made in constructing nuclear interactions from chiral effective field theory (EFT) \cite{epelbaum:05,entem:03a}, which is based on the most general local Lagrangian with nucleon and pion fields and all possible interactions consistent with the (broken) chiral 
symmetry of QCD. In contrast to phenomenological interaction models, the EFT approach is universal and provides a model-independent framework with a systematic organization of consistent $2N$, $3N$  and higher-body forces (and other operators) prescribed by the power counting.

For both phenomenological and EFT potentials,  nuclear few- and many-body calculations are complicated by strong short-range repulsion 
and tensor forces that necessitate highly correlated trial wave functions, non-perturbative resummations, 
and slowly convergent basis expansions.  However, the non-perturbative nature of inter-nucleon interactions is strongly scale dependent and can be radically softened by using the renormalization group (RG)  to lower the momentum cutoff that is present in all nuclear interactions.  A consequence is that many-body calculations become much more tractable at lower resolutions, resulting in calculations that are amenable to straightforward perturbative methods, simple variational ans\"atze, and rapidly convergent basis expansions~\cite{bogner:07b,bogner:07c, bogner:07a, bogner:07d}. The RG approach has the important advantage of being able to vary the cutoff as a tool to optimize and probe the quality of the many-body solution, and to provide estimates of omitted terms in the Hamiltonian.

The above considerations have motivated the construction of low-momentum 
potentials \vlowk through the renormalization group (RG) method
\cite{bogner:03a,bogner:03b}  and, more recently, by the similarity 
renormalization group (SRG) method \cite{bogner:07b,bogner:07c}. Both methods serve to
eliminate the strong coupling between low- and high-momentum modes in the Hamiltonian 
such that low-energy observables are preserved. In the RG method, one integrates out the problematic 
high-momentum components of the input interaction above a momentum cutoff $\Lambda$, leading to a new 
energy independent potential \vlowk that has the same low-energy on-shell transition
matrix (t-matrix) as the input potential. In the original approach $\Lambda$ constitutes a sharp cutoff above 
which the  t-matrix is zero;  the method has since been generalized to include a smooth momentum-space 
regulator to avoid technical difficulties stemming from the sharp cutoff \cite{bogner:07a}. On the other hand, 
the SRG method uses a continuous sequence of unitary transformations that weakens off-diagonal matrix elements, 
driving the Hamiltonian towards a band-diagonal form~\cite{bogner:07b, bogner:07c}. In contrast to the RG method,
SRG preserves both  low- and high-energy observables independent of the value of the flow parameter $\lambda$ that 
provides a measure of the spread of off-diagonal strength. However, as with the standard RG, the calculation of 
low-energy observables is decoupled from the high-momentum physics with SRG-evolved potentials 
(i.e., one can truncate intermediate state summations to low momenta without distorting low-energy observables).

Observables are scale-independent quantities. It is well-known that RG (SRG) transformations generate short-range many-body forces (in principle, up to $A$-body) that ``run'' with the cutoff to maintain exact $\Lambda$ ($\lambda$) independence of $A$-body observables. If the RG transformation is truncated at the $2N$ level, then the resulting cutoff-dependence in $3N$ observables may provide an estimate of omitted short-range $3N$ forces in the Hamiltonian. Along these lines, low-momentum $2N$ potentials have been recently used in three- and four-nucleon$(4N)$ bound state 
calculations  \cite{nogga:04a} as a means to assess the size of omitted higher-body 
forces by varying the cutoff. There, it was found that the induced $3N$ forces due to the truncation to low momentum are of the same order as the so-called ``bare'' $3N$ forces attributed to integrating out excitations of nucleons. That is, the cutoff-dependence of the $3N$ binding energies was rather weak, varying by only 1 MeV over a large cutoff range, which is comparable to the $0.7$-$1$ MeV binding provided by the missing ``bare'' $3N$ forces  in conventional models and EFT calculations. In this sense, the RG evolution to low momentum does not induce strong short-ranged three-body force contributions to these $3N$ bound state observables. Similar results were obtained in $4N$ bound state calculations, where the various $2N$ \vlowk calculations did not differ any more from the phenomenological Tjon-line than did calculations using $2N$ plus adjusted $3N$ forces.

In the current study, we extend the cutoff-dependence study of Ref.~\cite{nogga:04a} to $3N$ and $4N$ 
scattering observables. In particular, we apply RG- and SRG-evolved $2N$ interactions to study how the 
neutron-deuteron $(\nd)$ elastic vector analyzing power $A_y$ and the space star cross section in 
$\nd$ breakup change with the cutoff. These being the two major long-standing failures of realistic 
interactions in their description of $3N$ data at low energy, one would like to use cutoff-dependence 
as a tool to assess the sensitivity of these observables to omitted short-range $3N$ force effects. 
Likewise, the same applies to observables in $4N$ scattering that show large deviations to data, 
namely the total neutron-triton $(n\text{-}t)$ cross section $\sigma_t$ around
the resonance  region at neutron lab energy $E_n = 3.5$ MeV 
and the $\pHe$ \,\, $A_y$ that also misses the data by as much as 25 - 40\%.

In Section~\ref{sec:3N} we study $3N$ observables and 
in Section~\ref{sec:4N} $4N$ observables. 
Finally in Section~\ref{sec:concl} we present the conclusions.

\section{Three-nucleon observables}\label{sec:3N}

\begin{figure}[!]
\begin{center}
\includegraphics[scale=0.54]{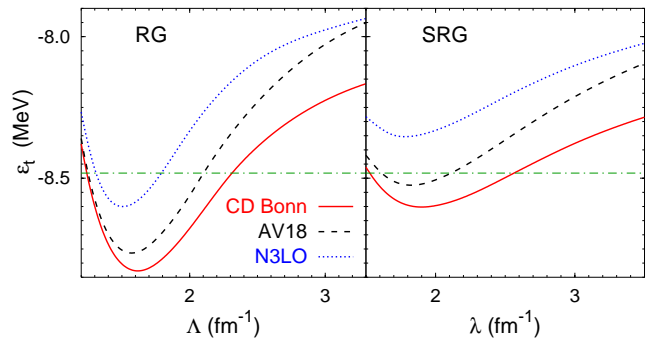}
\end{center} 
\caption{ \label{fig:eb} (Color online)
Triton binding energy as function of RG cutoff $\Lambda$ (left side)
and SRG parameter $\lambda$ (right side).
Results derived from CD Bonn (solid curves), AV18 (dashed curves),
and N3LO (dotted curves) potentials are shown.
The horizontal line at $\epsilon_t = -8.482$ MeV is the experimental value.}
\end{figure}

The results shown in this section are obtained from the solution of the 
symmetrized Alt, Grassberger and Sandhas (AGS) equations \cite{alt:67a} for 
the $3N$ system using the numerical techniques of 
\Ref~\cite{deltuva:03a}. In order to relate the present work 
to the findings of \Ref~\cite{nogga:04a} we repeat in \Fig~\ref{fig:eb} 
the cutoff dependence of the triton binding energy $\epsilon_t$
for CD-Bonn, AV18 and EFT potential at next-to-next-to-next-to leading order
(N3LO) \cite{entem:03a} based \vlowk
potentials using RG  (left side) and SRG (right side) methodologies. 
In contrast to the calculations of \Ref~\cite{nogga:04a}  with a sharp cutoff 
$\Lambda$, for simpler numerics we use a smooth regulator of
the form  $\exp{(-(k^2/\Lambda^2)^8)}$. The results are consistent with the
ones of \Refs~\cite{nogga:04a,bogner:07b}.
At first glance, the SRG parameter $\lambda$ that provides a measure of 
the spread of off-diagonal strength is not obviously related to the cutoff $\Lambda$ 
in the RG. However, in Ref.~\cite{bogner:07c} it was found that the ``decoupling scale'' 
for SRG-evolved interactions was of order $\lambda$. That is, low-energy phase shifts and 
binding energies are not distorted if high-momentum modes greater than the decoupling scale 
are set to zero (or any arbitrary value) by hand. Therefore, it is not surprising that the 
behavior of $\epsilon_t$ in terms of $\Lambda$ or $\lambda$ is qualitatively quite similar. 
We emphasize that the existence of cutoffs where $\epsilon_t$ agrees 
with the experimental value does \emph{not} imply vanishing $3N$ forces, as they will 
contribute to other observables.


\begin{figure}[!]
\begin{center}
\includegraphics[scale=0.54]{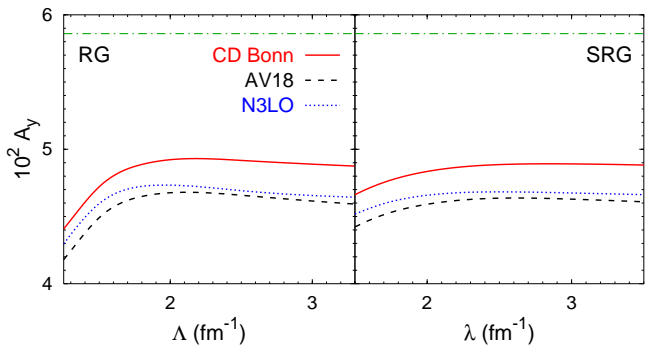}
\end{center} 
\caption{ \label{fig:nday} (Color online)
Neutron analyzing power $A_y$ for $\nd$ scattering at $E_n = 3$ MeV
and $\theta_{\cm} = 104$ deg as function of RG cutoff $\Lambda$ (left side)
 and SRG parameter $\lambda$ (right side). The 
 horizontal line at $10^2 \, A_y = 5.86$ is the
experimental value from \Ref~\cite{mcaninch:93}.}
\end{figure}

The neutron analyzing power $A_y$ in $\nd$ elastic scattering at 
neutron lab energy $E_n = 3$ MeV has a maximum at the center of mass 
(c.m.) scattering angle $\theta_{\cm} = 104$ deg, where
the predictions based on realistic interaction models underestimate
the experimental value by about 20 \%.
In \Fig~\ref{fig:nday} we plot the maximum value of $A_y$ as a function
of RG cutoff $\Lambda$  and SRG parameter $\lambda$. The cutoff dependence
is quite weak, indicating that this observable is not a sensitive probe of short-range
force effects. The net variation of $A_y$ over the range of cutoffs is smaller than the  
discrepancy from experiment of the initial interactions, which implies that short-range $3N$ forces
are not likely to solve the $A_y$ problem.


\begin{figure}[!]
\begin{center}
\includegraphics[scale=0.54]{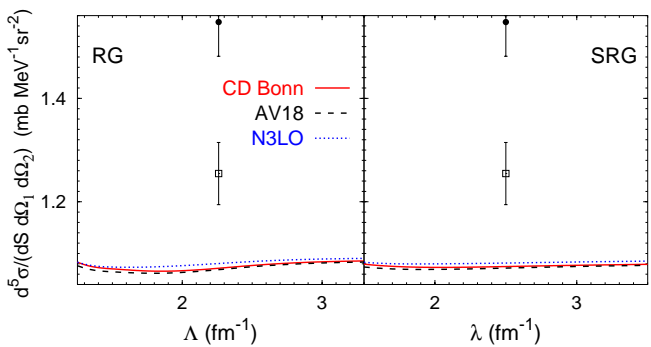}
\end{center} 
\caption{ \label{fig:ndss} (Color online)
Differential cross section for $\nd$ breakup at $E_n = 13$ MeV
in the space star configuration $(50.5^{\circ},50.5^{\circ},120^{\circ})$ 
at arclength $S=6.25$ MeV as function of RG cutoff $\Lambda$ (left side)
and SRG parameter $\lambda$ (right side). 
The experimental data are from \Ref~\cite{strate:89} (square)
and \cite{setze:05a} (circle).}
\end{figure}

The cutoff dependence is even weaker for the $\nd$ breakup differential cross section
in the space star configuration. We demonstrate that in \Fig~\ref{fig:ndss}
for the differential cross section
close to the center of the space star configuration at $E_n = 13$ MeV;
the values measured in two different experiments are shown as a reference.
These flat curves are again an indication that space star cross section is not
sensitive to short-range physics as already found in conventional calculations
with different $2N$ interactions or by adding a $3N$ force 
\cite{kuros:02b,deltuva:05a}.

\section{Four-Nucleon Observables}\label{sec:4N}

The results shown in this section are based on the solution of the 
AGS equations \cite{grassberger:67} in a symmetrized form following the technical 
developments expressed in \Ref~\cite{deltuva:07a,deltuva:07b,deltuva:07c} for 
all elastic and transfer $4N$ reactions below three-body breakup threshold.

As discussed in \Ref~\cite{deltuva:07a}, one of the simplest observables in 
$4N$ scattering is the total $\nH$ cross section $\sigma_t$ that 
exhibits a resonance around $E_n \simeq 3.5$ MeV. This peak of the 
total cross section results from a complicated interference between 
${}^3P_J \; \nH$ partial waves whose relative strength is 
sensitive to the realistic $2N$ force one uses. While at threshold we 
find the usual scaling between $\sigma_t$ and $\epsilon_t$ 
($\sigma_t$ decreases as $|\epsilon_t|$ increases), at $E_n \simeq 3.5$ MeV 
we observe a breakdown of scaling when we use N3LO \cite{deltuva:07a} 
which is a low-momentum potential when compared with the meson-exchange 
potentials. There N3LO yields the largest cross section 
while not having the lowest $|\epsilon_t|$. 

Furthermore, in \Ref~\cite{lazauskas:05a} it was found that adding the 
Urbana IX $3N$ 
force to AV18 slightly reduces $\sigma_t$ at the peak while more
significantly lowering the cross section at threshold towards the data 
as expected through scaling.

\begin{figure}[!]
\begin{center}
\includegraphics[scale=0.54]{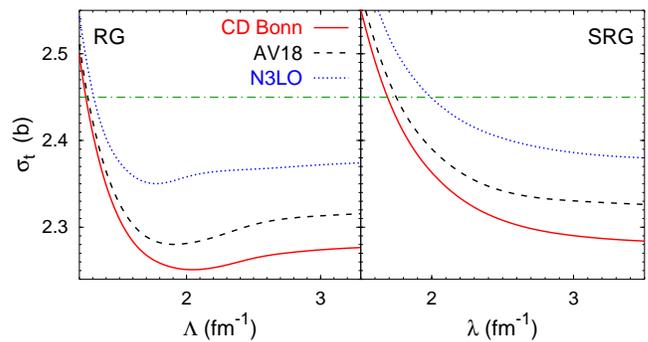}
\end{center} 
\caption{ \label{fig:nttot} (Color online)
Total cross section for $\nH$ scattering at $E_n = 3.5$ MeV
as function of RG cutoff $\Lambda$ (left side)
and SRG parameter $\lambda$ (right side).
The horizontal line at $\sigma_t = 2.45$ b is the
experimental value from \Ref~\cite{phillips:80}.}
\end{figure}

Therefore, in order to investigate the effect of low-momentum potentials on 
$\sigma_t$ we plot in \Fig~\ref{fig:nttot} the total cross section at the peak 
versus $\Lambda \, (\lambda)$. In contrast to studied $3N$ observables,
$\sigma_t$ shows stronger dependence on $\Lambda$ or $\lambda$,
which is not surprising since the ratio of triples to pairs increases. 

\begin{figure}[!]
\begin{center}
\includegraphics[scale=0.54]{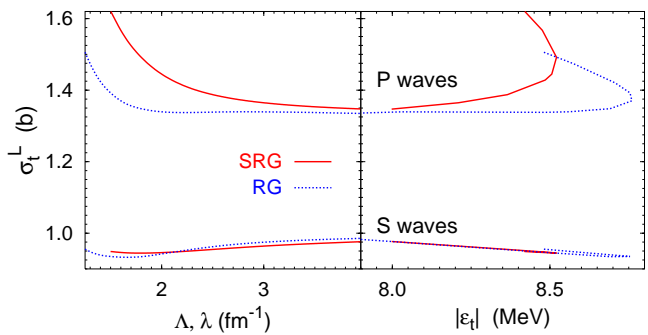}
\end{center} 
\caption{ \label{fig:ntSP} (Color online)
$S$-  and $P$-wave contributions to the total cross section for $\nH$ 
scattering at $E_n = 3.5$ MeV. On the left side they are shown as
functions of RG cutoff $\Lambda$ or SRG parameter $\lambda$,
while on the right side their correlation with the $\Hh$ binding energy is shown.
The SRG and RG interactions are derived from the AV18 potential.}
\end{figure}

\begin{figure}[!]
\begin{center}
\includegraphics[scale=0.58]{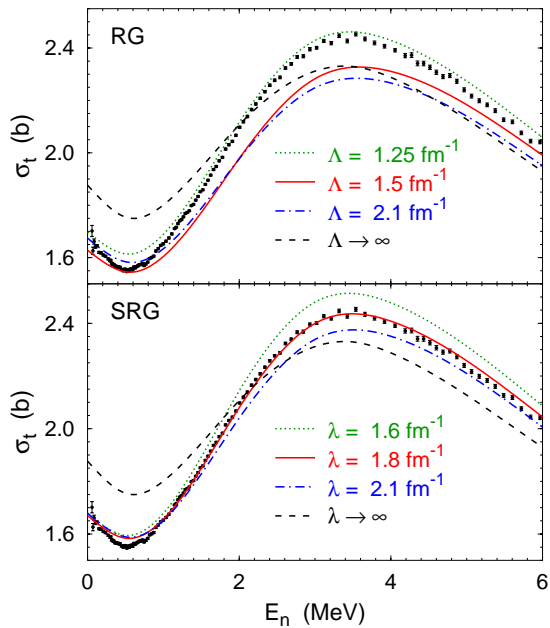}
\end{center} 
\caption{ \label{fig:nttotAV} (Color online)
Total cross section for $\nH$ scattering 
as function of neutron lab energy for different values of
RG cutoff $\Lambda$ (top) and SRG parameter $\lambda$ (bottom).
All results are derived from the AV18 potential.
The predictions of the original AV18 potential (dashed curves) are also shown.
The experimental data are from \Ref~\cite{phillips:80}.}
\end{figure}

In \Fig~\ref{fig:ntSP} we split up the total cross section into $\nH$ relative
$S$-  and $P$-wave contributions using  AV18-based $V_{\mathrm{low}\: k}$.
The $S$-wave contribution scales well with the $\Hh$ binding energy; 
that scaling is slightly violated for RG approach at $\Lambda < 1.5 \fm^{-1}$.
In contrast, $P$-waves show no correlation with $\epsilon_t$ and 
are responsible for an increase of the total cross section at small
$\Lambda \, (\lambda)$ values. This is consistent with the findings 
of \Ref~\cite{deltuva:07a}.
 In \Fig~\ref{fig:nttotAV} we use the AV18 potential to show $\sigma_t$ versus 
$E_n$ for the values of  $\Lambda \, (\lambda)$ that fit the experimental 
triton  binding energy and for the one that yields deepest binding.
While one finds that one may describe the total neutron cross section over a 
wide energy range by using $\Lambda \approx 1.25 \fm^{-1}$ in RG method or 
$\lambda \approx 1.8 \fm^{-1}$ in the SRG approach that  also yield
reasonable values for $\epsilon_t$, we emphasize once again
that these particular values of $\Lambda \, (\lambda)$  do not imply vanishing 
$3N$ and $4N$ forces, as they will contribute to other few- and many-nucleon observables,
for example, to the ground state energies of light nuclei that do not match experiment
with those ``special'' choices of $\Lambda \, (\lambda)$ \cite{bogner:07d}.

\begin{figure}[!]
\begin{center}
\includegraphics[scale=0.54]{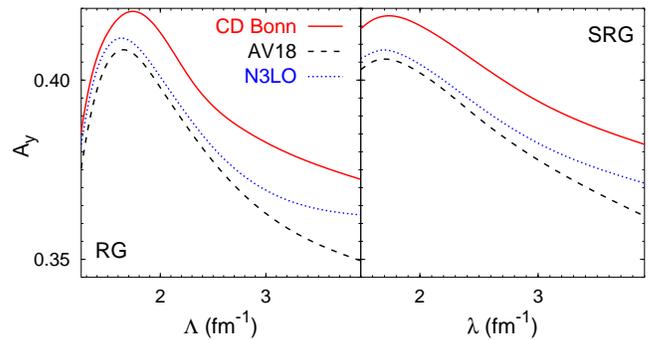}
\end{center} 
\caption{ \label{fig:ntay} (Color online)
Maximum of the 
neutron analyzing power $A_y$ for $\nH$ scattering at $E_n = 3.5$ MeV
as function of RG cutoff $\Lambda$ (left side)
 and SRG parameter $\lambda$ (right side).}
\end{figure}

\begin{figure}[!]
\begin{center}
\includegraphics[scale=0.54]{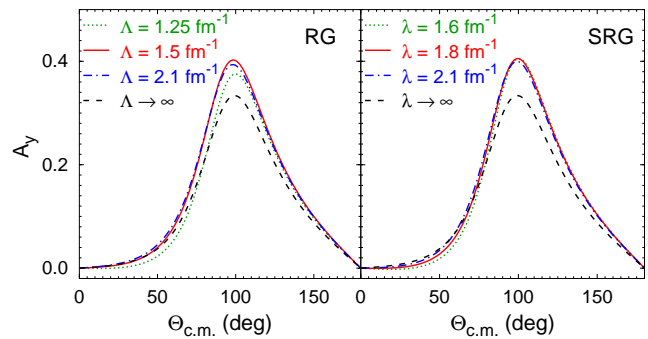}
\end{center} 
\caption{ \label{fig:ntAVay} (Color online)
Neutron analyzing power $A_y$ for $\nH$ scattering at $E_n = 3.5$ MeV
as function of c.m. scattering angle for different values of
RG cutoff $\Lambda$ (left side) and SRG parameter $\lambda$ (right side).
All results are derived from the AV18 potential.
The predictions of the original AV18 potential (dashed curves) are also shown.}
\end{figure}

There is a clear correlation between maximum values
of  nucleon analyzing power $A_y$ in $\pHe$ and $\nH$
scattering \cite{deltuva:07a,deltuva:07b}; we therefore study only
the latter case. Though $A_y$ in $\nd$ and $\nH$ scattering are also
correlated to some extent, their dependence on cutoff is different
as shown in \Fig~\ref{fig:ntay}; it is considerably stronger for $\nH$.
 The largest increase of $A_y$ value
at the maximum, by a factor 1.13 (N3LO) to 1.21 (AV18),
is observed around cutoff values that yield experimental or deepest binding.
However, according to \Ref~\cite{deltuva:07b}, the experimental $A_y$ value 
at the maximum for $\pHe$ scattering in the same energy region
is larger than theoretical predictions by a factor 1.45 (CD Bonn) to 1.55 (AV18).
In \Fig~\ref{fig:ntAVay} we use the AV18 potential to show $A_y$ versus $\theta_{\cm}$
for the values of  $\Lambda \, (\lambda)$ that fit the experimental 
triton  binding energy and for the one that yields deepest binding.

\section{Conclusions}\label{sec:concl}

In order to probe the sensitivity of $3N$ and $4N$ scattering observables to short-range physics, 
we used AV18, CD Bonn and N3LO based \vlowk potentials that are generated through 
the RG (SRG) method to study their evolution with the cutoff $\Lambda \, (\lambda)$. Truncating 
the RG (SRG) equations to the two-body level amounts to neglecting
short-ranged $3N$ (and higher) forces that are generated to preserve exact cutoff independence.
Therefore, one expects to find residual cutoff dependence in few-body observables when only $2N$ 
low momentum interactions are used. That cutoff dependence may provide a measure
of the sensitivity of a given observable to omitted short-ranged $3N$ (and higher) forces 
since the RG evolution does not distort the long-ranged forces arising from pion
exchange, provided $\Lambda$ (or $\lambda$) is well above the pion mass, and it is only
short-ranged operators that ``run'' to maintain cutoff independence.

Comparing the results shown in \Figs~\ref{fig:nday},\ref{fig:ndss} with those in 
\Figs~\ref{fig:nttot},\ref{fig:ntay} one cannot help noticing that the cutoff dependence of 
$3N$ observables is much weaker than the one observed for $4N$ observables. Clearly for the $3N$ 
observables, the cutoff dependence is rather weak, which seems to imply that short-ranged 
$3N$ forces are not likely to fix the two long-standing discrepancies with data mentioned above. 
This is indeed what has been found when the leading missing ``bare" $3N$ force, which contains
both long- and short-ranged operators, is added. 
Nucleon-deuteron  $A_y$ in elastic scattering and the space star differential cross section for breakup
barely change by adding a two-$\pi$-exchange $3N$ force \cite{witala:01,kievsky:01a,kuros:02b},
or an effective $3N$ force due to the explicit $\Delta$-isobar excitation \cite{deltuva:03c,deltuva:05a}, 
or the more recent leading $3N$ force from chiral EFT \cite{epelbaum:02a}. 
However, there is hope that the subleading long-range $3N$ forces from chiral EFT might be important 
for the resolution of these problems due to their novel space, spin, and isospin structures. 

On the contrary $4N$ scattering observables seem to be more sensitive to omitted short-ranged 
many-body forces as demonstrated by the more pronounced dependence on the cutoff. In $\nH$ scattering 
at low energy the total cross section $\sigma_t$ is dominated  by $S$ and $P$ waves in the relative 
$\nH$ motion. The $S$ waves $({}^1S_0, {}^3S_1)$ are Pauli repulsive 
and therefore simply scale with $\epsilon_t$ over the whole energy region shown in 
\Fig~\ref{fig:nttotAV}. Therefore sensitivity to $2N$ forces comes through the $P$ waves 
$({}^3P_0, {}^3P_1 - {}^1P_1, {}^3P_2)$ which, in the resonance region, have a very complex 
behavior with the cutoff parameter, leading to breaking of scaling with  $\epsilon_t$. 
This is consistent with the previous findings \cite{deltuva:07a} obtained with various $2N$ potentials.
It also indicates that the $\sigma_t$ discrepancy may be sensitive to missing short or intermediate 
range $3N$ forces, in contrast to the $\pHe$ $A_y$ puzzle \cite{deltuva:07b}.

From these studies one may conclude that $4N$ scattering observables are more 
sensitive to short range physics than the $3N$ observables where, at low energy, they seem to be 
constrained, to a large extent, by on-shell $2N$ scattering and three-particle unitarity, 
as was expressed long ago by  Brayshaw \cite{brayshaw:77}. Recent developments 
\cite{witala:01,kievsky:01a} indicate that one needs to fit triton binding energy or 
neutron-deuteron doublet scattering length to constrain some other $3N$ observables that, unlike $A_y$,
are sensitive to scaling. Nevertheless, this is already fine tuning on top of results that are already
very close to the experimental data. This is not the case for low-energy $4N$ observables.

\begin{acknowledgments}
A.D. is supported by the Funda\c{c}\~{a}o para a Ci\^{e}ncia e a Tecnologia
(FCT) grant SFRH/BPD/34628/2007 and A.C.F. in part by the FCT grant POCTI/ISFL/2/275.
\end{acknowledgments}


\end{document}